\documentclass[10pt,a4paper,twocolumn]{article}
\usepackage{amsmath}
\usepackage{amssymb}
\usepackage[pdftex]{color,graphicx}
\usepackage{yfonts}
\usepackage{xargs}
\usepackage{url}
\usepackage{bm}
\usepackage[pdftex,dvipsnames]{xcolor}  % Coloured text etc.
\usepackage[noadjust]{cite}  
\usepackage{changes} 
\usepackage{algorithm}
\usepackage{algpseudocode}
\usepackage{tikz}
\usepackage{hyperref}
\usepackage[a4paper, total={6.9in, 9in}]{geometry}

\definecolor{orange}{cmyk}{0,0.42,1,0}

\newcommand\smalld{
  \mathchoice
    {{\scriptstyle\mathcal{D}}}% \displaystyle
    {{\scriptstyle\mathcal{D}}}% \textstyle
    {{\scriptscriptstyle\mathcal{D}}}% \scriptstyle
    {\scalebox{.7}{$\scriptscriptstyle\mathcal{D}$}}%\scriptscriptstyle
 }
  
\newcommand\smalls{
  \mathchoice
    {{\scriptstyle\mathcal{S}}}% \displaystyle
    {{\scriptstyle\mathcal{S}}}% \textstyle
    {{\scriptscriptstyle\mathcal{S}}}% \scriptstyle
    {\scalebox{.7}{$\scriptscriptstyle\mathcal{O}$}}%\scriptscriptstyle
 }

\newcommand\copyrighttext{%
  \footnotesize \textcopyright 2016 IEEE. Personal use of this material is permitted.
  Permission from IEEE must be obtained for all other uses, in any current or future 
  media, including reprinting/republishing this material for advertising or promotional 
  purposes, creating new collective works, for resale or redistribution to servers or 
  lists, or reuse of any copyrighted component of this work in other works. \\
  DOI: \href{http://dx.doi.org/10.1109/TNSRE.2018.2794184}{10.1109/TNSRE.2018.2794184}}

\newcommand\copyrightnotice{%
\begin{tikzpicture}[remember picture,overlay]
\node[anchor=south,yshift=20pt] at (current page.south) {\fbox{\parbox{\dimexpr\textwidth-\fboxsep-\fboxrule\relax}{\copyrighttext}}};
\end{tikzpicture}%
}

\begin{document}
\title{Surrogate-based artifact removal from single-channel EEG}
\date{}
\author{M. Chavez$^{1,2}$, F. Grosselin$^{1,3}$, A. Bussalb$^{1,2}$, F. De Vico Fallani$^{1,2, 4}$, and X. Navarro-Sune$^{1,2,5}$ \\ %\vspace{-5px}%
	\small
	$^1$CNRS, UMR-7225, Paris, France\\ \vspace{-1px}
	\small
	$^2$Sorbonne Universit\'{e}s, UPMC Univ Paris 06, France \\ \vspace{-1px}
	 \small
	$^3$myBrain Technologies. Paris, France\\ \vspace{-1px}
	 \small
	$^4$INRIA Paris, ARAMIS team, Paris, France\\ \vspace{-1px}
	\small
	$^5$INSERM UMRS-1158 Neurophysiologie Respiratoire Exp\'{e}rimentale et Clinique, Paris, France \\ \vspace{-1px}
}

\maketitle
\copyrightnotice
%\vspace{-1cm}

\begin{abstract}
\vspace{-0.3cm}
\emph{Objective:} The recent emergence and success of electroencephalography (EEG) in low-cost portable devices, has opened the door to a new generation of applications processing a small number of EEG channels for health monitoring and brain-computer interfacing. These recordings are, however, contaminated by many sources of noise degrading the signals of interest, thus compromising the interpretation of the underlying brain state. In this work, we propose a new data-driven algorithm to effectively remove  ocular and muscular artifacts from single-channel EEG: the surrogate-based artifact removal (SuBAR). \emph{Methods:} By means of the time-frequency analysis of surrogate data, our approach is able to identify and filter automatically ocular and muscular artifacts embedded in single-channel EEG. \emph{Results:} In a comparative study using artificially contaminated EEG signals, the efficacy of the algorithm in terms of noise removal and signal distortion was superior to other traditionally-employed single-channel EEG denoising techniques: wavelet thresholding and the canonical correlation analysis combined with an advanced version of the empirical mode decomposition. Even in the presence of mild and severe artifacts, our artifact removal method provides a relative error 4 to 5 times lower than traditional techniques. \emph{Significance:} In view of these results, the SuBAR method is a promising solution for mobile environments, such as ambulatory healthcare systems, sleep stage scoring or anesthesia monitoring, where very few EEG channels or even a single channel is available.

\end{abstract}

\section{Introduction}
\vspace{-0.3cm}
Electroencephalogram (EEG) s the standard recording of electrophysiological activity of the brain. Due to its temporal resolution (ms), technical simplicity (portable and non-invasive) and low cost,  EEG is nowadays extensively used for studying different cognitive and pathological brain states. EEG recordings are, however, often contaminated by non-neural physiological activities, as well as other external or environmental noises, that seriously degrade the signals of interest. 

Eye movement-related artifacts have a strong detrimental effect on the quality of scalp EEG. During eye movements, abrupt changes on the retina's resting potential are primarily observed in the frontal EEG electrodes before their widespread propagation over the scalp. The strength and spatial distribution of the artifact strongly depends on the position of EEG electrodes and on the direction of the eye movement~\cite{sweeney2012a, sweeney2012b, uriguen2015}. Eye blinks also contaminate  EEG signals but with an artifact whose amplitude is generally larger than that produced by eye movements.  Muscular artifacts originate from the electrical activity elicited by contracting muscles. Although electromyographic (EMG) activity can be observed over the entire scalp, the amplitude and distribution of EMG artifacts depend on the type of muscle contracted (e.g. jaw, neck or face) and on the degree of tension~\cite{sweeney2012a, sweeney2012b, uriguen2015}. Other possible perturbations include breathing artifacts, electrodermal interferences produced by sweating, motion artifacts (e.g. head or chest movements), as well as shifts of the electrical properties of electrodes. 

In clinical settings, visual inspection and manual removal of contaminated EEG segments is a common practice prior to any off-line signal analysis. Obviously, such manual methods  are not suitable for on-line applications. There is a number of general techniques used for artifact removal from EEG recordings. When the frequency bands of the signal and interferences do not overlap,  simple low pass, band pass or high pass filtering are effective techniques for removing artifacts. Nevertheless, some interferences (e.g. muscular activities) have a wide spectral distribution that overlaps with that of EEG, making difficult to remove them. In case of spectral overlap, more refined techniques such as adaptive filtering, Wiener filtering, as well as blind source separation (BSS) methods have been effectively used to cancel EEG interferences. Other  methods like wavelet decompositions (WT) and empirical mode decomposition (EMD) have also been successfully applied to remove EEG artifacts~\cite{patel2014, khatun2016}. For a review and discussion of different approaches see~\cite{sweeney2012a, sweeney2012b, uriguen2015} and references therein.
 
Linear regressions (in time or frequency domain) and adaptive filtering have been successfully applied in EOG and ECG correction procedures~\cite{uriguen2015}. Their main  disadvantage is, however, that they assume that one or more reference channels with the artifacts waveforms are available. Independent component analysis (ICA) is a blind signal separation (BSS) technique that has been largely used for EEG artifact removal~\cite{iriarte2003, delorme2007}. Briefly, ICA separates multi-channel EEG signals into statistically independent components which are assumed to represent the underlying sources of the observed EEG signals. Clean EEG signals can be reconstructed  by removing artifact-related components from the original ICA decomposition. ICA-based methods can effectively remove interferences from a wide variety of artifactual sources in EEG recordings only when the number of channels and the amount of data are large enough~\cite{uriguen2015}. Canonical correlation analysis (CCA) is a more efficient BSS method for muscular artifact removal that exploits the relative low autocorrelation of EMG artifacts in comparison with EEG activity~\cite{declercq2006, sweeney2013}.

Artifact removal by wavelet-based methods and other data-driven decompositions (e.g. the EMD and its variants~\cite{huang1998, colominas2014, uriguen2015}) relies on the assumption that EEG artifacts can be represented by one or more levels or modes, which are thresholded before reconstructing the signal from the filtered representation. The success of these techniques depends on the threshold selecting criteria. In recent studies, different combinations of algorithms to remove artifacts (e.g. EMD-ICA, EMD-CCA) have provided significantly improved results~\cite{sweeney2013, chen2014, chen2016}.

  % FIGURE 1
\begin{figure*}[ht!] 
   \centering%from left, bottom, right and top
   \includegraphics[width=0.95\linewidth]{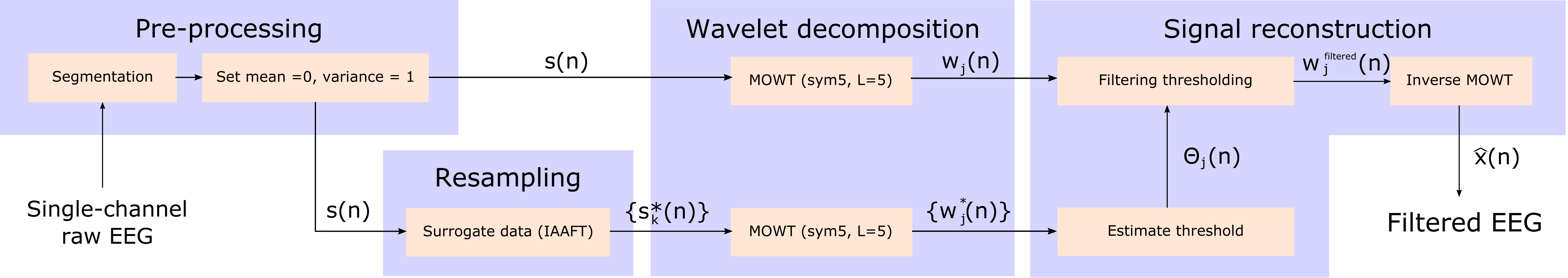} 
   \caption{Block diagram of the proposed artifact removal method.}
   \label{fig:pipeline}
\end{figure*} 

In last years, simplified EEG systems with few channels have been developed with the aim to increase usability of ambulatory neuroimaging technologies in clinical environments (e.g. in epilepsy and sleep diagnosis) and in custom-designed settings for routine monitoring~\cite{casson2010, mihajlovic2015}. For some applications like neurofeedback, mental state classification,  emotion sensing, etc., artifact removal algorithms should perform reasonably well with short epochs of streaming EEG data. Further, conventional multichannel techniques can not be applied directly to isolate artifact sources in reduced channels configurations. Hence, there is a growing need to develop effective artifact removal techniques that can operate in short segments of single channel EEG, especially in single-channel settings~\cite{sweeney2012a,chen2014, chen2016, khatun2016}. 

In this work, we propose a data-driven approach for artifact removal from single-channel EEG: the Surrogates-Based Artifact Removal (SuBAR) method. Our approach combines wavelet decomposition and a resampling method called surrogate data. Under the hypothesis of stationary EEG segments, muscular and ocular artifacts are considered as nonstationary events that can be identified across different scales. The obtained scales from the original recorded EEG are then compared with those obtained from resampled data, which are stationary by construction. Hence, instead of estimating the filtering threshold from uncontaminated segments or theoretical functions, we obtain it directly from the WT of surrogate data.  

The proposed framework is validated on artifact-free EEG data contaminated with simulated artifacts of several types (EMG or EOG). The method is also illustrated on a real EEG contaminated data collected from a healthy subject. The reliability and performances of our method are also compared with those obtained by an unsupervised wavelet-based artifact removal~\cite{sweeney2013, chen2016, uriguen2015, khatun2016} and by those resulting from the CCA in combination with an advanced version of the  EMD~\cite{sweeney2013, colominas2014, chen2016}. The remainder of the paper is organized as follows: Section~\ref{meth} describes the proposed framework, as well as the WT and CCA-EMD approaches. Section~\ref{sect:database} presents the database and Section~\ref{sect:evaluation} describes the procedure to simulate EMG and EOG artifacts~\cite{delorme2007}. Section~\ref{results} provides the experimental results and evaluation of the method. Finally, we conclude the paper with a discussion in Section~\ref{conclusion}.

\section{Methods}
\label{meth}
In this section, we describe the various techniques employed in this work. Firstly, we describe our surrogates-based EEG technique, and then we outline two alternative methods for artifact removal from EEG single-channels (used here for comparison): a wavelet thresholding method~\cite{sweeney2013, chen2016, uriguen2015, khatun2016}, and the artifact removal algorithm based on CCA combined with the EMD~\cite{sweeney2013}.

\subsection{Resampling-based artifact removal}
In this work, we assume that a recorded single-channel EEG signal\footnote{In this paper, scalars are denoted by italic letters, vectors by lowercase bold-face letters and matrices by uppercase, bold-face leterrs. Superscript $^T$ refers to the transpose operation.} $ \mathbf{s}= \left[s(1), \ldots,s(N)\right]^T$  is a linear combination of the original desired stationary EEG signal $\mathbf{x}= \left[x(1), \ldots,x(N)\right]^T$ contaminated with an artifact $\mathbf{v}= \left[v(1), \ldots,v(N)\right]^T$ plus instrumental Gaussian white noise $\boldsymbol{ \eta}$. Although some nonlinear filters have been proposed to remove multiplicative EMG interferences~\cite{chen2010}, here we only consider additive artifacts of the form $\mathbf{s} = \mathbf{x} + \mathbf{v}+\boldsymbol{\eta}$ for the sake of simplicity. 

The aim of our work is to filter $\mathbf{v}$ from the vector of observations $\mathbf{s}$, with minimal \emph{a priori} knowledge on the artifact signal, the EEG signal and the observational noise. To this end, we have used wavelet transform as a tool to detect and remove artifacts from single EEG channels. Wavelet-based artifact removal aims at separating the artifact components from the clean EEG components in the wavelet domain. Once the artifact components have been identified and the corresponding wavelet coefficients removed, the remaining components are kept to reconstruct the cleaned EEG signal. The thresholding criterion is traditionally based on statistical properties of the wavelet spectrum obtained from uncontaminated baselines or from theoretical thresholding functions. The originality of our approach is that we propose to estimate this threshold directly from the stationarized spectrum of the observed data.

Figure~\ref{fig:pipeline} shows the general pipeline of the artifact removal method proposed in this study.  The key point of our method is that the hypothesis of EEG stationarity (which corresponds to time-invariance in the time-frequency spectrum) is statistically characterized on the basis of a set of surrogates which all share the same average stationary spectrum as the desired EEG signal. In our approach, the recorded EEG signal is of the form $\mathbf{s} = \mathbf{x} + \mathbf{v}+\boldsymbol{\eta}$, where the artifact $\mathbf{v}$ is assumed to be a non-stationary event restricted to a finite time interval shorter than the observed signal $\mathbf{ s}$. Since surrogates can be viewed as distinct, independent stationary realizations of the observed signal, the wavelet  spectra obtained from surrogates can define the learning set for stationarity. Wavelet thresholding is therefore based on the statistical distribution of the spectrum of surrogates (for each time-frequency bin). All the components of the original wavelet decomposition higher than a given threshold  are set to zero, and the desired EEG signal is reconstructed utilizing the remaining components only.

The main blocks of the proposed method are the following: 

\subsubsection*{\textbf{Pre-processing}} In most EEG systems, some procedures are currently applied to prepare the EEG data for analysis. These procedures include re-referencing of recorded signals, down-sampling of original data for saving transmission power and computational cost, or filtering for baseline and power-line (50/60 Hz) interference removal. In our pre-processing step, the sampled raw EEG signal is divided into epochs of size $N$. Very short segments ($<1$~s) may not represent some slow artifacts properly (e.g. ocular or movement artifacts) whereas in very large windows (e.g. $>5$~s), the stationary assumption of EEG may be no longer valid and there is a high chance that clean EEG segments will be distorted by the artifact removal procedure~\cite{islam2016}. Here, we set $N = 3.5$~s.

\subsubsection*{\textbf{Resampling}} Bootstrap and other resampling methods have been used extensively in the past to appropriately determine the properties of a time series before applying different analysis and modeling techniques~\cite{efronBOOK}. The aim of these techniques is to capture a given structure of the original signal, and construct additional realizations that can then be used to test the original data for additional, unexplained structure. Surrogate data techniques were proposed as non-parametric methods for testing general hypotheses on data without making assumptions on the underlying generating process. Surrogates are time series created directly from the original dataset through replication of the linear autocorrelation (or equivalently, the power spectrum density) and amplitude distribution, with all other higher-order quantities randomized~\cite{schreiber2000}. 

In this work, surrogate time series $\mathbf{s}^*_k\;; k=1, \ldots,K$ were obtained by destroying the organized phase structure that controls the nonstationarity of the original signal $\mathbf{s}$~\cite{theiler1996}. In the classical Fourier-transform surrogate algorithm (FT), the signal is first Fourier transformed, and the magnitude of the spectrum is then kept unchanged while its phases are replaced by a random sequence, uniformly distributed over $\left[-\pi, \pi\right]$~\cite{theiler1996, schreiber2000}. This modified spectrum is then inverse Fourier transformed, leading to a Gaussian stationary surrogate time series with the same spectrum as the original signal~\cite{theiler1996, schreiber2000}. To deal with non-Gaussian data, the algorithm of amplitude adjusted Fourier transform (AAFT) first orders a Gaussian white noise time series to match the rank order of the original data and derives the FT surrogate of this time series~\cite{theiler1996}. The final surrogate is scaled to the distribution of the original data by sorting the original data according to the ranking of the FT surrogate.~\cite{theiler1996}. The AAFT algorithm guarantees that surrogate data $\mathbf{s}^*_k$ possesses the original distribution exactly and the original power spectrum approximately~\cite{schreiber2000}.  In this work, we have used the so-called Iterative Amplitude Adjusted Fourier Transform (IAAFT), which is an iterative version of AAFT~\cite{schreiber1996, schreiber2000}. The steps are repeated until the autocorrelation function is sufficiently similar to the original, or until there is no change in the amplitudes~\cite{schreiber1996, schreiber2000}. 

In the context of nonlinear time series analysis, surrogate data have been widely used for testing linearity and, more recently, for testing stationarity~\cite{schreiber2000, borgnat2010}. To this end, time-frequency decompositions are used for determining whether the spectral characteristics of a signal change significantly over time, which is an indication of non-stationarity~\cite{borgnat2010}. The time-frequency localization properties of the wavelet transform allow a comparison of the original signal with the \emph{stationarized} surrogates in the wavelet domain. We can therefore detect the position, on the time-frequency plane, where the spectrum of the observed EEG differs from a stationary process. In this study, values derived from $K=1000$ surrogate realizations were used to represent the spectrum of stationary signals. 

\subsubsection*{\textbf{Wavelet decomposition}} Wavelet transform is a method commonly used  to remove artifacts from EEG signals~\cite{sweeney2013, patel2014, chen2016, uriguen2015, khatun2016}. The wavelet transform is an useful analysis tool for time-frequency representation of non-stationary signals, obtained by convolving the signal $\mathbf{s}$ with a scaled and translated wavelet function, 
\begin{equation}
\psi_{a,b}(n) = \frac{1}{\sqrt{a}}\psi\left ( \frac{n-b}{a}\right) ,
\end{equation}
with $a$ and $b$, positive real numbers and $n = 1, \ldots, N$. The WT provides thus a decomposition of the signal in different scales, where the obtained wavelet coefficients represent a measure of similarity between the signal $\mathbf{s}$ and the corresponding wavelet function $\psi_{a,b}(n)$. 

In the discrete wavelet transform (DWT) both factors $a$ and $b$ are integers and are chosen in a dyadic grid ($a=2^j$ and $b = k2^j$ with integers $j$ and $k$ playing roles of the decomposition level and temporal localization at this level, respectively). At the  $j$th level of decomposition, a  matrix $\mathbf{M}_j$ of size $N \times N$ can be appropriately built with the corresponding orthonormal wavelet basis. The wavelet coefficients $\bm{w}_j =\left[w_j(1), w_j(2),\ldots,w_j(N)\right]^T$ can be thus obtained by $\bm{w}_j = \mathbf{M}_j\mathbf{s}$. The DWT can be seen as a band-pass filter bank: the decomposition starts by passing the signal through a low-pass filter giving the approximation coefficients, and a high-pass filter producing the detail coefficients. The resulting two orthogonal sub-bands are afterwards down-sampled by two. Then, the low-pass result can be recursively filtered by the same pair of filters until the desired frequency range is obtained. 

In this paper, we used the Maximal Overlap Wavelet Transform (MODWT)~\cite{percivalBOOK}, instead of the orthogonal DWT, to estimate wavelet decomposition. As the DWT, the MODWT can also be utilized for multi-resolution analysis, with the advantage that it can be applied to signals of any size, while the DWT requires the sample size $N$ to be an integer power of two. In contrast with the usual DWT, the MODWT is translation invariant, i.e. shifting circularly the signal by any amount results into shifting the outputs of the low-pass and high-pass filters by the same amount. This property does not hold for the DWT because of the subsampling involved in the filtering process~\cite{percivalBOOK}. 

Let $\{ h_{j,l}\; ;  l = 0,...,L_j\}$ and $\{g_{j,l}\; ;  l = 0,...,L_j\}$ a $j$th level wavelet and scaling filter, respectively. The corresponding $j$th level MODWT wavelet and scaling filters are defined, respectively, by $\{\widetilde{h}_{j,l} = h_{j,l}/2^{j/2}\}$  and  $\{\widetilde{g}_{j,l} = g_{j,l}/2^{j/2}\}$ with the same common length $L_j$~\cite{percivalBOOK}. The $j$th level MODWT wavelet and scaling coefficients are $N$ dimensional vectors defined by 
\begin{equation}
w_{j}(n)=\sum_{l=0}^{L_j-1}\widetilde{h}_{j,l}s(n-l \mod N)
\end{equation}
and 
\begin{equation}
v_{j}(n)=\sum_{l=0}^{L_j-1}\widetilde{g}_{j,l}s(n-l \mod N) \; ,  
\end{equation}
for $n = 1,\ldots,N-1$, respectively. The ``mod'' operator denotes the modular arithmetic between two integers. 

Inverse transforming the MODWT coefficients creates the so-called details $\bm{\smalld}_j$ and smooths $\bm{\smalls}_j$ that form a multi-resolution analysis of signal $\mathbf{s}$~\cite{percivalBOOK}: 
\begin{equation}
\mathbf{s} = \bm{\smalls}_J + \sum_{j=1}^J \bm{\smalld}_j \; ,
\end{equation}
with $J$ denoting the number of decomposition levels. 
An interesting property of MODWT is that $\bm{\smalld}_j$ and $\bm{\smalls}_j$ are associated with zero phase filters~\cite{percivalBOOK}. The frequency band of detail coefficients associated to coefficients $\bm{w}_j$  is given by $2^{-(j+1)}\leqslant f \leqslant 2^{-j}$ with a width $\omega_j = 2^{-(j+1)}$~\cite{percivalBOOK}. In contrast with DWT, there are always $N$ coefficients at each level. The MODWT is an energy preserving transform and the total energy of $\mathbf{s}$ can be partitioned by the MODWT scaling and wavelet coefficients: $\|\mathbf{s}\|^2 = \sum_{j=1}^{J}\|\bm{w}_j\|^2 + \|\bm{v}_J\|^2$~\cite{percivalBOOK}. 

For all the wavelet-based methods studied in this paper, we used the symlets wavelet and 5 levels of decomposition. The symlets are orthogonal functions, nearly symmetrical wavelets with an oscillatory waveform and good time-frequency localization properties~\cite{daubechiesBOOK}. This makes it suitable wavelet choice for filtering and reconstructing EEG signals~\cite{aniyan2014, alkadi2012}.  

\subsubsection*{\textbf{Signal reconstruction}} Comparison of the original signal with the stationary surrogates in the wavelet domain can identify non-stationary events on the time-frequency plane. The wavelet coefficients corresponding to artifacts are expected to be of high amplitude and well localized in time and scales, while the clean EEG coefficients are expected to be small and homogeneously spread over the whole segment.

To detect artifact components, the wavelet transform of surrogates are firstly averaged to produce a mean reference spectrum of stationarity:
\begin{equation}
\label{eq:mediaSurrogates}
\overline{\bm{w^*_j}}= \frac{1}{K}\sum_{k=1}^{K}\bm{w}_j^k \; ,
\end{equation}
where $K$ is the number of surrogates and $\bm{w}_j^k$ denotes the wavelet coefficients for surrogate $k$ at the $j$th level. At each time-point of the $j$th level of decomposition, the standard deviation for the ensemble of surrogates  can also be determined as
\begin{equation}
\label{eq:stdSurrogates}
\sigma^*_j(n) = \sqrt{\frac{1}{K}\sum_{k=1}^K \left(w_j^k(n) - \overline{w^*_j(n)}\right)^2}\; .
\end{equation}

At each point of the wavelet domain, the significance of wavelet coefficients at a given level was assessed by quantifying its statistical deviation from values obtained for the ensemble of surrogates.
Thus, non-stationary components in observed signals can be detected by comparing $\bm{w}_j^k$ values to a given threshold. 

Distribution of wavelet coefficients from surrogates can be fitted with an appropriate distribution, such as Gaussian or Gamma distribution, and then setting a one-sided confidence interval, for rejection of non-stationary components. Here, the significance was obtained by the ratio $\Lambda_j(n)=(w_j(n) - \overline{w^*_j(n)})/\sigma^*_j(n)$ whose p-value is given by the Chebyshev's inequality: for \emph{any} statistical distribution of $w_j(n)$: $p(| \Lambda_j(n) | \geqslant \Theta_j (n)) \leqslant 1/\Theta_j ^{2}(n)$ where $\Theta_j(n) $ is the chosen statistical threshold~\cite{papoulisBOOK}.  

The threshold values are compared with the wavelet coefficients of the original signal in the following manner:
\begin{equation}
\label{eq:filtering}
w^{\mathrm{filtered}}_j(n) =
  \begin{cases}
    \overline{w^*_j(n)} & \text{if }  w_j(n)  \geqslant  \Theta_j(n)   \\
    w_j(n) &  \text{otherwise}
  \end{cases},
\end{equation}

If the original wavelet coefficients are greater than the threshold, they are set to the average value of the reference spectrum (obtained from the surrogates). In this manner, only the stationary components of the original spectrum are retained. Here, the threshold $\Theta_j(n)$ was set as the values 
greater than $95$\% of the values in the surrogate distribution.
After non-stationary components have been filtered, the cleaned signal can be recomposed from all levels using the inverse MODWT~\cite{percivalBOOK} of the cleaned coefficients $\bm{w}^{\mathrm{filtered}}_j$.

The main steps of proposed SuBAR method for automatic artifact removal are summarized in Algorithm~\ref{algo1}.

\begin{algorithm}
\caption{Surrogate-based artifact removal (SuBAR) algorithm for single-channel EEG}\label{algo1}
\textbf{Input:} Signal $\mathbf{s}$,  number of surrogates $K$ (e.g. $K=1000$), threshold $\alpha$ (e.g. $5\%$ of significance level obtained from the surrogate distribution), number of  decomposition levels $J$\\
\textbf{Output:} Filtered signal $\mathbf{s}^{\mathrm{filtered}}$ 
\hrule
\begin{algorithmic}[1]
\State Estimate the wavelet coefficients (MODWT) $\bm{w}_j$ from input signal $\mathbf{s}$
\For{each $ k=0 \ldots$ $K$ }
    \State Create a surrogate time series $\mathbf{s}^*_k$ from signal $\mathbf{s}$
    \State Estimate the wavelet coefficients (MOWDT) $\bm{w}_j^k$ from surrogate $k$
\EndFor
\State Compare the wavelet coefficients of the original signal with those obtained from surrogates
 \For{each level of decomposition $j=1 \ldots  J$ }
      \State Estimate the threshold $\Theta_j(n)$ from the surrogate distribution and the significance level $\alpha$
      \If{$w_j(n)\geqslant \Theta_j(n)$}
        \State $w^{\mathrm{filtered}}_j(n) = \overline{w^*_j(n)}$ 
        \Else
        \State $w^{\mathrm{filtered}}_j(n) = w_j(n)$
        \EndIf
 \EndFor
	 \State Use the cleaned coefficients $\bm{w}^{\mathrm{filtered}}_j$ to reconstruct the filtered signal $\mathbf{s}^{\mathrm{filtered}}$ with the inverse MODWT
\State \textbf{return} $\mathbf{s}^{\mathrm{filtered}}$
\end{algorithmic}
\end{algorithm}

\subsection{Methods for Comparison}
The proposed technique is compared with two other commonly used methods for artifact removal from single-channel EEG signals: \textit{i)} a wavelet-based artifact removal~\cite{sweeney2013, chen2016, uriguen2015, khatun2016} which is based on the classical wavelet-thresholding and \textit{ii)} a single-channel method based on the combination of CCA with the EMD~\cite{sweeney2013, colominas2014, chen2016}.

\subsubsection{\textbf{Wavelet thresholding}} The key point of the wavelet thresholding is to separate, in the wavelet domain, the artifact components from the uncontaminated EEG components. For thresholding the wavelet coefficients we have used a level-dependent threshold~\cite{donoho1995, johnstone97, islam2016}: $\Theta_j^{\mathrm{wthr}} =\sigma_j \sqrt{2 \ln N}$, where $N$ is the length of signal and $\sigma_j^2$ is the estimated noise variance for the wavelet coefficients, $\bm{w}_j$, at the $j$th level of decomposition, which is usually calculated by~\cite{johnstone97}: $\sigma_j^2 = \mathrm{median}\left( |\bm{w}_j|/0.6745\right)$. Such level-dependent thresholding is more appropriate than a single universal threshold in case of correlated and non-Gaussian data, as such that it characterizes the underlying EEG activities~\cite{johnstone97}. Artifact removal is finally obtained by removing wavelet coefficients whose absolute values exceed the threshold~\cite{safieddine2012, uriguen2015, khatun2016} as follows:  
\begin{equation}
\label{eq:filteringWT}
w^{\mathrm{filtered}}_j(n) =
  \begin{cases}
    0 & \text{if }  w_j(n)  \geqslant  \Theta_j(n)^{\mathrm{wthr}}  \\
    w_j(n) &  \text{otherwise}
  \end{cases},
\end{equation}

\subsubsection{\textbf{Canonical Correlation Analysis (CCA) combined with Empirical Mode Decomposition for single channels}}
\label{cca}
CCA is a BSS technique currently used for separating a number of mixed or contaminated  signals~\cite{declercq2006}. The recorded multichannel EEG signals are considered as a mixture given by $\mathbf{S} = \mathbf{A}\mathbf{S_o}$,where $\mathbf{A}$ is the unknown mixing matrix and the components in $\mathbf{S_o}$ are the statistically independent and unknown source signals, which include the artifacts. As other BSS techniques, CCA estimates the mixing matrix and recovers the original sources as $\widehat{\mathbf{S_o}}=\mathbf{V}\mathbf{S}$, where $\mathbf{V}$ is the unmixing matrix, i.e. the estimate of the inverse of $\mathbf{A}$. Artifacts are removed by computing $\widehat{\mathbf{X}} = \mathbf{A}_\mathrm{clean}\widehat{\mathbf{S_o}}$, where $\mathbf{A}_\mathrm{clean}$ is the mixing matrix with the columns corresponding to artifact related sources, set to zero.

Due to their relatively low autocorrelation, muscle artifacts are generally well identified in the last components obtained by the CCA algorithm~\cite{declercq2006}. Previous studies have shown that CCA outperforms different ICA algorithms for artifact removal from multichannel EEG and fMRI signals~\cite{borga2002, declercq2006, gao2010}. Other advantages of the CCA include that, i) as CCA uses second order statistics, it is a more computationally efficient algorithm than ICA and, ii) contrary to ICA algorithms,  the CCA method always provides the same result for a given input.

As ICA, the original CCA is a multi-variate technique that requires multi-channel recordings to perform the decomposition. In some recent works, multi-dimensional time series have been generated from a single-channel recording, using popular data-driven methods such as the wavelet transform and the empirical mode decomposition~\cite{sweeney2012a, sweeney2012b, sweeney2013, uriguen2015, chen2016}. Artifact removal is then performed by applying BSS techniques (e.g. ICA or CCA) to the generated multi-channel signals. For our single-channel artifact removal technique we have here combined the CCA, as the best choice form multi-channel artifact removal, with an improved version of the EMD  as the best choice for time series decomposition~\cite{sweeney2013}. The combination of these methods has been shown to be more reliable and computationally more efficient than the EMD combined with ICA~\cite{sweeney2013}. 

% FIGURE 2
\begin{figure}[h!] 
   \centering%from left, bottom, right and top
   \includegraphics[width=0.95\linewidth]{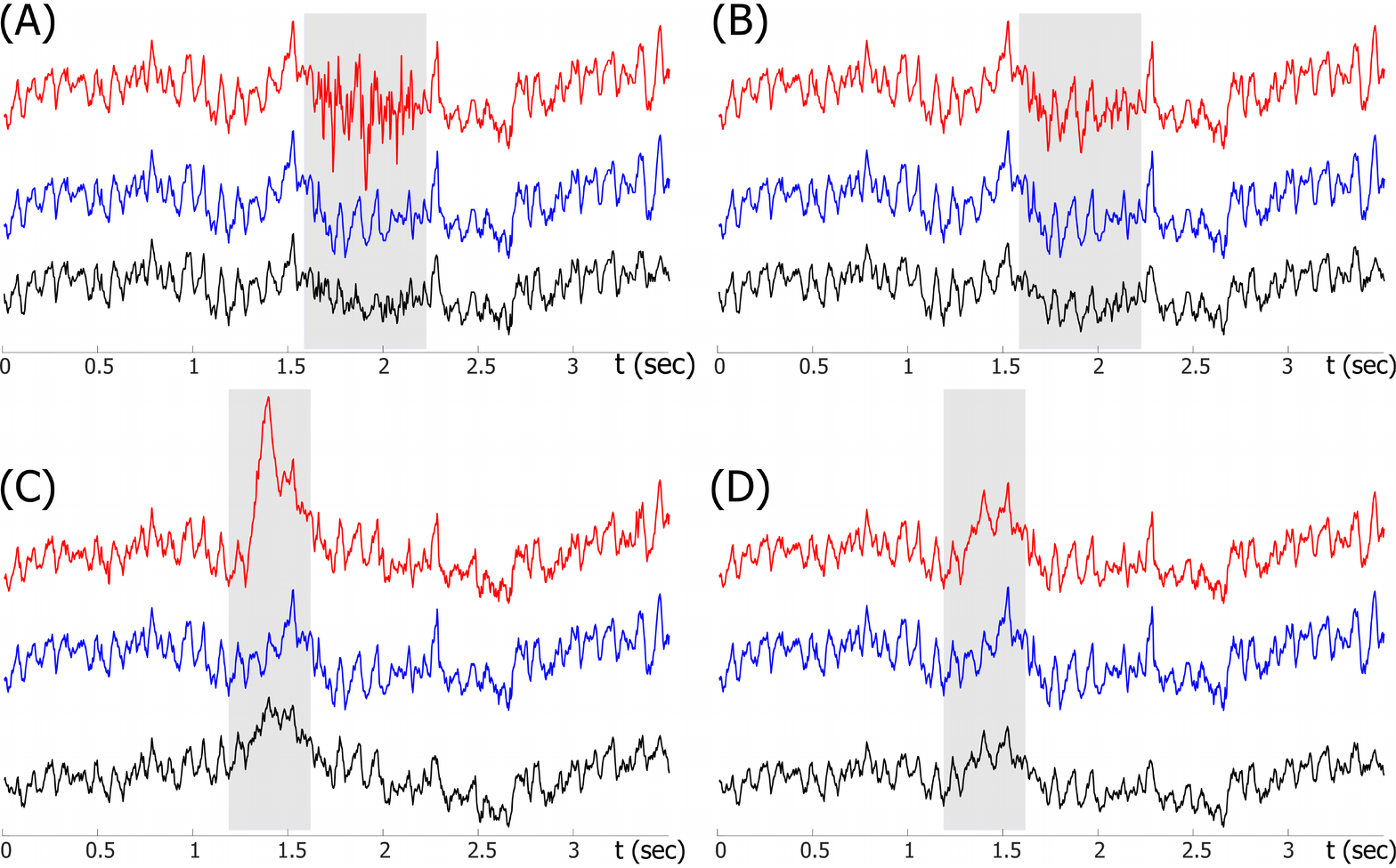} 
   \caption{Examples of original clean EEG signals (blue curves), with superimposed artifacts (red curves) and removal results after the proposed algorithm (black curves). The EEG signals are contaminated by:  simulated large (A) and small (B) muscular activities, superimposed large (C) and small (D) ocular artifact. Gray boxes indicate the artifactual regions.}
   \label{fig:artifacts}
\end{figure} 

The orignal EMD is an adaptive data-driven method, proposed by~\cite{huang1998}, to decompose non-linear and non-stationary signals into a number of sub-components called intrinsic modal functions (IMFs), with well defined instantaneous frequencies. The original signal is thus decomposed as $\mathbf{s}= \mathbf{r} + \sum_{i} \mathbf{c}_{i}$, where $\mathbf{r}$ stands for a residual trend, and the
intrinsic modes $\mathbf{c}_{i}$'s are nearly orthogonal to each other~\cite{huang1998}. By construction, the spectral supports are decreased when going from one residual to the next. Nevertheless their frequency discrimination applies only locally (in time) and they cannot correspond to a sub-band filtering~\cite{flandrin2004}. Despite its several advantages to decompose mixed signals, the EMD of noisy data may result in a corruption of modes, i.e. very similar oscillations appear in different IMFs~\cite{huang1998}. Recently, the so-called complete Ensemble EMD with adaptive noise (CEEMDAN) was proposed to ameliorate the spectral separation of modes and to reduce computational time~\cite{torres2011, colominas2014}. The key idea on this algorithm relies on averaging the modes obtained by EMD applied to several realizations of Gaussian white noise added to the original signal. 

Here, we use this decomposition technique (CEEMDAN), to convert the single-channel signal $\mathbf{s}$ into a multi-channel signal $\mathbf{S}$. By means of the CCA, the source signals associated to artifacts can be then removed as described before. The cleaned single-channel signal without the artifacts can be finally reconstructed by adding the new IMFs components in $\widehat{\mathbf{X}}$~\cite{sweeney2013}. Hereafter, for the sake of simplicity, we denote this technique CCA-EMD.

\subsubsection*{Criteria for artifact removal with the CCA-EMD method}
Eye blinks artifacts display large slow waves and have large autocorrelation compared to EEG sources. Here, EOG artifacts were thus identified from the first canonical variates due to their large autocorrelations (larger than $0.9$)~\cite{sweeney2013}. In contrast, due to the frequency spectrum of the EMG artifacts, they resemble high frequency activity. In this work, the CCA components with spectral bandwidth larger to $15$~Hz were associated to muscle artifacts and removed from the reconstruction~\cite{declercq2006, delorme2007, Muthukumaraswamy2013}. Other filtering criteria can also be applied, possibly providing better tuning of the algorithm to the particularities of other EEG artifacts.

\section{Database}
\label{sect:database}
To assess the performance of the proposed artifact removal technique, we employed two datasets recorded via surface electrodes (Acticap, BrainProducts GmbH, Germany) using $N_e=$64 scalp positions according to the standard 10-10 montage.  The first dataset consisted on a collection of clean EEG signals from two subjects who were instructed to remain quietly, but alert, with their eyes closed during two minutes. The second dataset was composed by EEG signals (two minutes) from one subject instructed to deliberately produced artifacts by eye blinking and jaw clenching at short intervals. To verify the correct realization of artifacts and to detect their instances, we used four external EOG and EMG channels (right and left frontalis and anterior temporalis muscles). In all recordings, impedance between electrodes and skin were set below $5$~k$\Omega$. According to the declaration of Helsinki, written informed consent was obtained from subjects after explanation of the study, which was approved by the ethical committee CPP-IDF-VI of Paris (n\textsuperscript{o} 2016-A00626-45). 

The EEG signals were amplified, digitized at a sampling frequency of 1024 Hz, then down-sampled to 256 Hz and segmented in $3.5$ sec non-overlapped windows to reduce computational cost in successive blocks. Clean signals from the first dataset were artificially contaminated with muscle artifacts and eye movements as in~\cite{delorme2007} and were used to compare quantitatively the removal of artifacts by the different methods. Trained experts at the EEG platform of the ICM visually inspected all trials and selected different artifact-free EEG segments from the first database. Finally, contaminated signals (with vertical ocular blinks and other pronounced muscular artifacts) were selected from the second dataset to qualitatively illustrate the efficacy of the denoising process. 

% FIGURE 3
\begin{figure}[h!] 
   \centering%from left, bottom, right and top
   \includegraphics[width=0.9\linewidth]{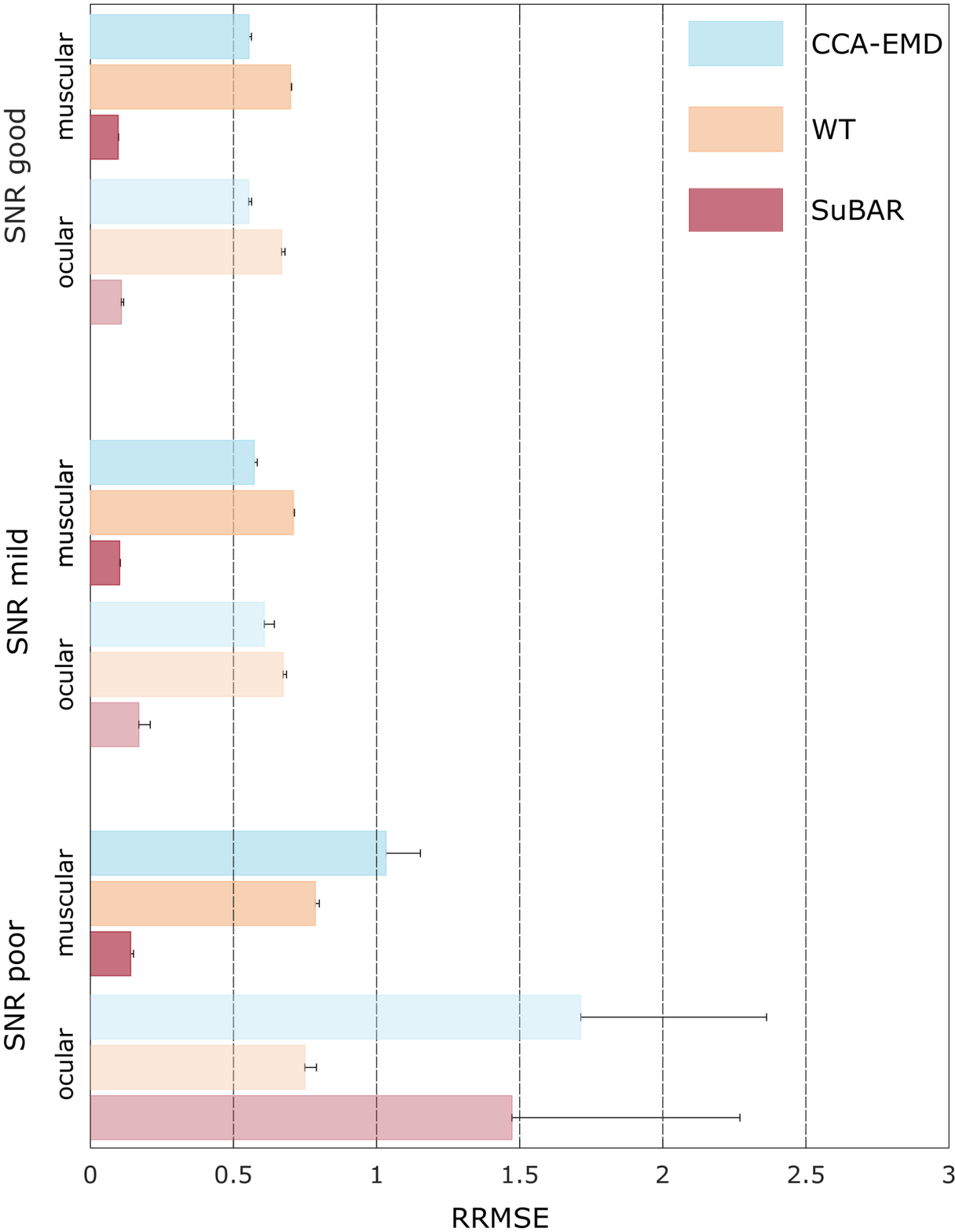} 
   \caption{General comparison of artifact removal as a function of SNR. The bars represent average of all signals and all EEG channels. Error bars represent standard error. Good SNRs range from 0 to 5 dB, mild SNRs range from -10 to 0 dB, and poor SNRs are those values less than 0 dB. The  algorithms are: Wavelet Thresholding(WT), the Canonical Correlation Analysis combined with Empirical Mode Decomposition (CCA-EMD), and  the proposed SuBAR method.}
   \label{fig:artifactsRRMSE}
\end{figure} 

\section{Evaluation}
\label{sect:evaluation}
Artificially contaminated EEG signals were simulated using clean segments from the first dataset (absence of ocular and muscular activity and other artifacts due to body movements or technical interferences). Artifacts, superimposed to clean data, were generated in three steps: 

% FIGURE 4
\begin{figure*}[th!] 
   \centering%from left, bottom, right and top
   \includegraphics[width=0.95\linewidth]{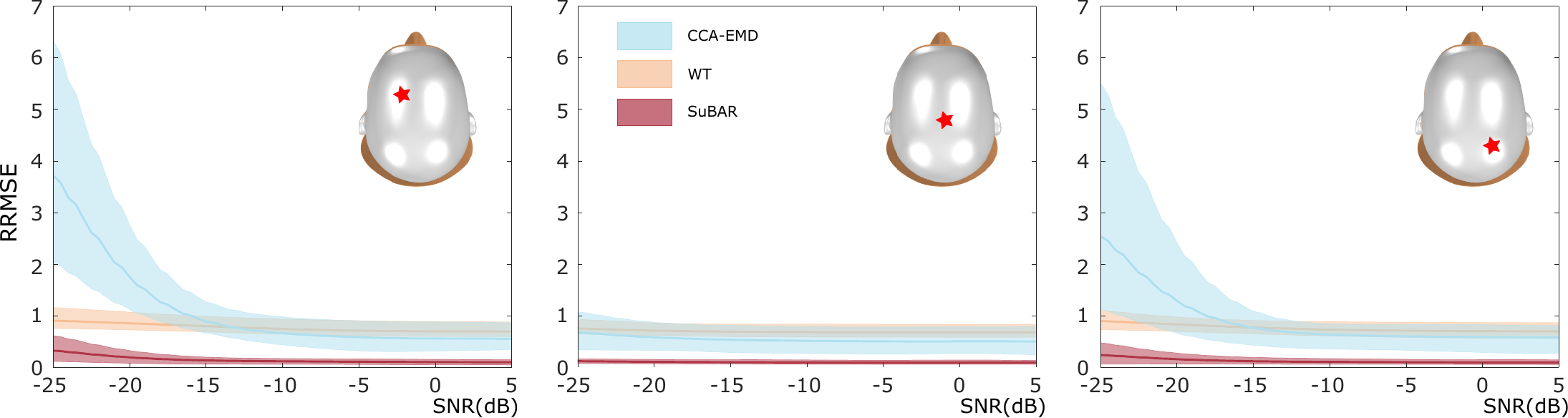} 
   \caption{Examples of RRMSE as a function of SNR on the EEG epochs containing simulated muscular artifacts. Red asterisks indicate the scalp position of electrodes (from left to right: FC1, Cz and CP2). Solid curves indicate mean values and shadowed areas display the $5$th and $95$th percentiles.}
   \label{fig:artifactsEMG}
\end{figure*} 

\subsection{Characterization of spatial distribution}
Since artifacts of different origins have a specific distribution in the scalp, we first computed a weight vector $ \mathbf{a_{art}}$ of dimensions $N_e \times 1$  to scale the artifact patterns according to the topographical information from the scalp electrodes. To obtain the topographical information for each type of artifact, we applied ICA decomposition to a selection of real EEG segments containing the artifacts. Vector $\mathbf{a_{art}}$ was the rescaled column in the estimated mixing matrix $\mathbf{A}$ associated to the artifactual component, found by inspecting some features such as the autocorrelation and spatial position in the scalp~\cite{delorme2007}. The components associated to eye movements and blinks ($\mathbf{a_{art}^{o}}$) were detected as those yielding high amplitudes on the most frontal electrodes, and muscular components  ($\mathbf{a_{art}^{m}}$)  as those having transient and fast activities localized most importantly on temporal electrodes.

\subsection{Generation of artifact patterns}
The two different artifact patterns consisted on vector signals $\mathbf{r}$ of length $N$  obtained as follows:
\begin{itemize}
\item Ocular artifacts, $\mathbf{r_{o}}$, were recorded from the original electrooculogram signal (EOG).
\item Muscular artifacts, $\mathbf{r_{m}}$, were generated using random noise band-pass filtered between $20$ and $60$~Hz with a random length between $0.3$-$0.8$~s (equivalent to those observed in real EEG data)~\cite{delorme2007}.
\end{itemize}
From the above patterns, the matrix of simulated artifacts, $\mathbf{V}$ of dimension $Ne \times N$, was computed by performing the product $\mathbf{a_{art}} \mathbf{r}$ for each pattern. 
Clean EEG signals (selected by visual inspection) and simulated artifacts came from different subjects to ensure that all segments (trials) of simulated and real EEG data were independent of each other. 

\subsection{Setting of signal to noise ratios}
Synthetic artifacts $\mathbf{V}$ were superimposed on the clean EEG segments $\mathbf{b}$ as follows: $\mathbf{b}^\mathrm{artifacted} = \mathbf{b} + \lambda\mathbf{v}$, where $\lambda$ represents the contribution of the artifact. For each trial and artifact type, the signal to noise ratio (SNR) was adjusted by changing the parameter $\lambda$ as follows: 
\begin{equation}
\label{eq:snr}
\mathrm{SNR} = \frac{\mathrm{RMS}(\overline{\mathbf{b}})}{\mathrm{RMS}(\mathbf{v})}
\end{equation}
where $\mathrm{RMS}(\overline{\mathbf{b}})$ corresponds to the  root mean squared value averaged over all channels, and $\mathrm{RMS}(\mathbf{v})$ denotes the root mean squared value of the artifact. Following~\cite{delorme2007}, prior to computing SNR for the topographic artifacts, we scaled their amplitudes by the highest channel gain in the applied scalp map. Here, the artifact contribution was gradually decreased in a dB scale ($10\log_{10}(\mathrm{RMS})$)  from $-25$ dB to $5$dB. 

The performances of the considered algorithms for artifact removal were evaluated both in terms of the amount of artifact reduction and the amount of distortion they bring into clean EEG signals. Performances were expressed in terms of the relative root mean square error (RRMSE)~\cite{wang2009}:
\begin{equation}
\label{eq:rrmse}
\mathrm{RRMSE} = \frac{\mathrm{RMS}(\widehat{\mathbf{x}}-\mathbf{x})}{\mathrm{RMS}(\mathbf{x})}
\end{equation}
where $\widehat{\mathbf{x}}$ is the signal after artifact removal. To assess whether our technique preserves the frequency spectrum of clean EEG, or it introduces any phase shift in denoised signals, we have also measured the spectral coherence and phase delay between the corrected data and the clean EEG segments.

\section{Results and Discussion}
\label{results}
The proposed algorithm is applied to  EEG data contaminated by  simulated muscular activities and ocular artifacts. Figure~\ref{fig:artifacts} shows some examples of added artifacts and their removal by the SuBAR algorithm. Results suggest a good removal of both types of artifacts without distorting the background EEG signals outside the artifactual regions. Low- or band-pass filters were not capable of removing muscular artifacts without altering the underlying brain activity because the overlap of the frequency spectrum of artifacts and that of clean EEG signals. 

We examined the reliability of the SuBAR method at different SNR values in terms of the RRMSE as mentioned in Section~\ref{sect:evaluation}.  We also applied two alternative methods -- wavelet thresholding and CCA combined with the advanced version of the EMD -- to the contaminated EEG channels, and performed a comparison through 20 independent simulations. In our simulations, the amplitude of artifacts were scaled by their spatial distribution and a prescribed SNR. Here, we present the performances of different methods for a reduced number of EEG channels which spatial positions are relevant for different practical applications  of ambulatory clinical neuroimaging, in reduced settings for routine monitoring~\cite{casson2010, mihajlovic2015} or for practical BCI systems based on motor imagery. EEG electrodes included are: FC1, FC2, FCz, CP1, CP2, CPz, C1,C2 and Cz.

% FIGURE 5
\begin{figure*}[ht!] 
   \centering%from left, bottom, right and top
   \includegraphics[width=0.95\linewidth]{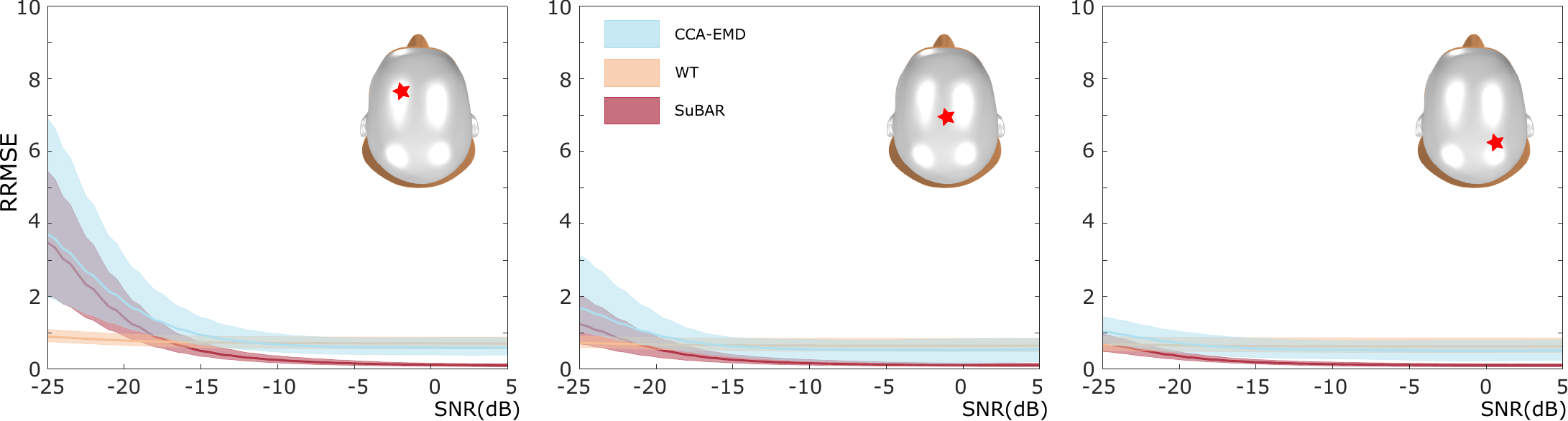} 
   \caption{The RRMSE as a function of SNR on the EEG epochs contaminated with ocular artifacts. Red asterisks indicate the scalp position of electrodes (from left to right: FC1, Cz and CP2). Same stipulations as in the caption of  Fig.~\ref{fig:artifactsEMG}}
   \label{fig:artifactsEOG}
\end{figure*}

Figure~\ref{fig:artifactsRRMSE} shows a systematic comparison of different algorithms, expressed in terms of the RRMSE for different SNR and both muscular and ocular artifacts. Bar plots display averaged values estimated over all channels and segments. Results clearly indicate that, even with severe artifacts, our SuBAR method yields better performances than traditional artifact removal techniques.

\subsubsection*{Muscular artifacts} The performances of different methods to remove simulated muscular artifacts are shown in Figure~\ref{fig:artifactsEMG}. It is clearly seen that the SuBAR method outperformed its competitors for all SNRs. Although wavelet thresholding performances are  stable for all SNRs, this technique is not able to recover the original EEG signals. On the other hand, as the contamination level increases in frontal and posterior regions, the performances of the data-driven CCA-EMD method are considerably degraded. Filtering modes from the CCA-EMD method is insufficient to remove large muscular artifacts without altering the underlying brain activity since the frequency spectrum of the muscle artifacts overlaps with that of the brain signals. 

Results support the hypothesis that, thanks to the time-frequency localization properties of the wavelet transform, a comparison of the contaminated EEG signal with the surrogates in the wavelet domain can identify artifacts as non-stationary events embedded on a stationary signal. Long and persistant muscular artifacts could not be detected as the spectrum of the contaminated EEG will not differ from a stationary process.

\subsubsection*{Ocular artifacts} Figure~\ref{fig:artifactsEOG} shows the RRMSE for the different ocular artifact removal algorithms as a function of SNR. It can be observed that both the surrogate-based algorithm and the CCA-EMD methods  are not able to remove large artifacts without distorting the true signals, whereas the wavelet thresholding is an algorithm that provides better performances. This indicates that surrogates of EEG signals with large ocular artifacts cannot be distinguished in the wavelet domain from the decomposition of contaminated signals. Nevertheless, for mild and weak artifacts, our surrogate-based technique outperforms other methods and can better remove the artifacts and recover the underlying EEG signals. 

\subsubsection*{Distortion of clean EEG segments} 
\begin{figure}[h!] 
   \centering%from left, bottom, right and top
   \includegraphics[width=0.7\linewidth]{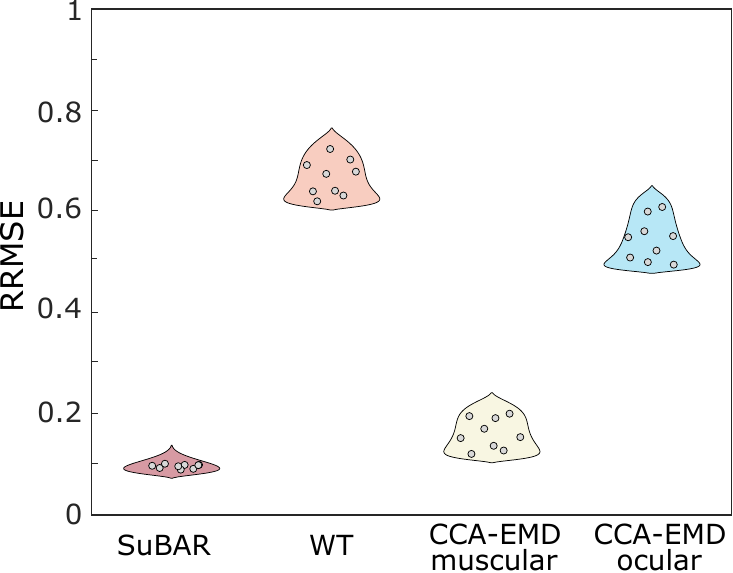} 
   \caption{The RRMSE of different methods when applied to artifact free EEG epochs (averaged over all segments for each channel). SuBAR is the proposed method, WT stands for wavelet thresholding, CCA-EMD muscular and ocular denote the criteria for the corresponding artifacts removal.}
   \label{fig:artifactsBIAS}
\end{figure}

Let us now quantify the distortions --in terms of RRMSE-- produced by each method when applied to artifact free EEG epochs. Figure~\ref{fig:artifactsBIAS} shows that, although wavelet thresholding is able to remove large ocular artifacts, it also produces an important amount of distortion on clean EEG segments. Similarly, the automatic correction of artifacts with CCA-EMD method altered clean EEG signals substantially, although the algorithm to detect ocular artifacts resulted in a larger distortion than the algorithm for muscular artifact removal. These results indicate that for single-EEG channels, the SuBAR method preserves better the EEG signals than the other considered artifact removal algorithms, which may remove true neural components from clean brain signals. 

\subsubsection*{Spectral distortion produced by the  SuBAR method} 
\begin{figure}[h!] 
   \centering%from left, bottom, right and top
   \includegraphics[width=\linewidth]{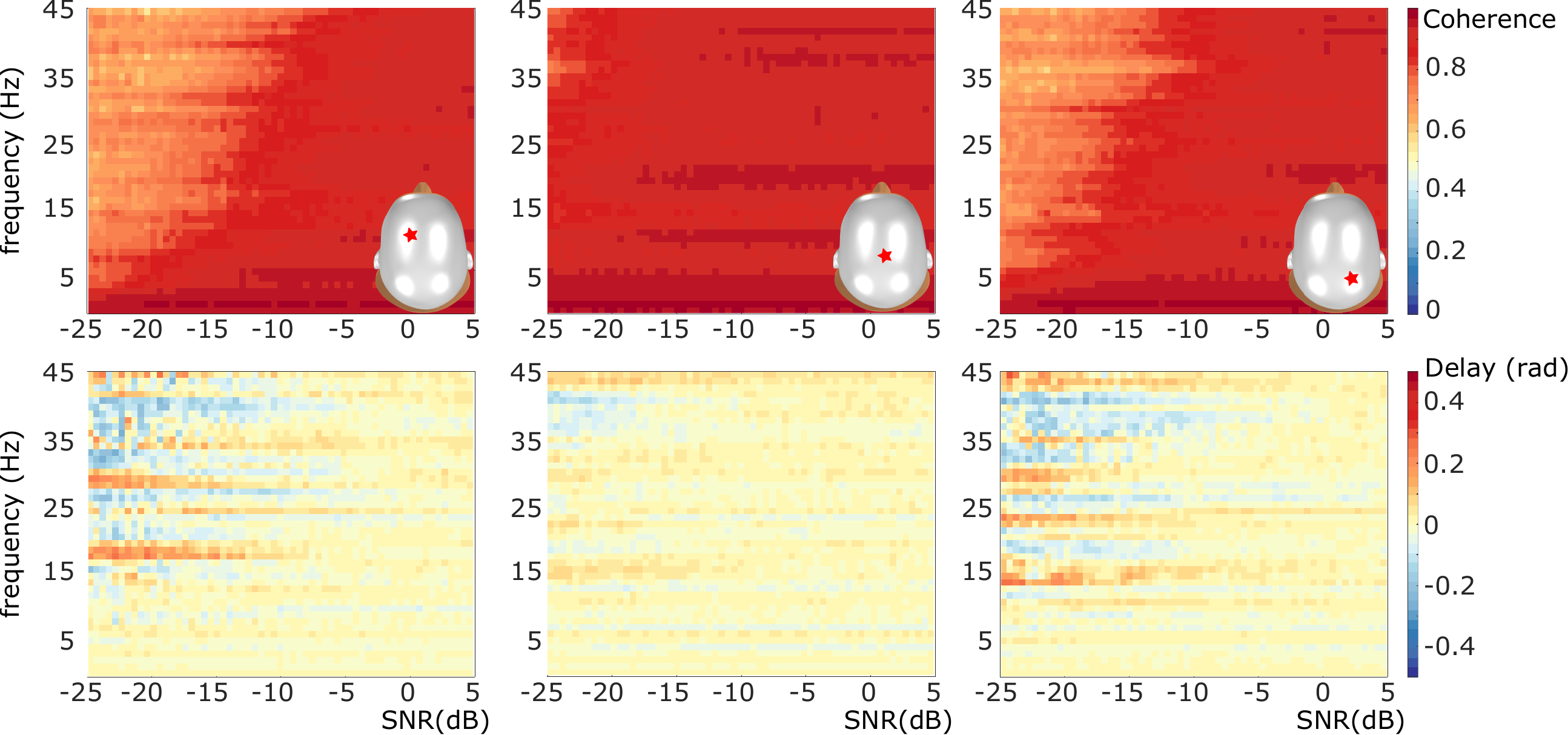} 
   \caption{Distortion produced by the SuBAR method on EEG epochs contaminated with muscular artifacts: The coherence (top plots) and phase delay (bottom plots) between original clean and denoised data as a function of SNR. Red asterisks indicate the scalp position of electrodes.}
   \label{fig:artifactsCOHERENCE_EMG}
\end{figure}

\begin{figure}[h!] 
   \centering%from left, bottom, right and top
   \includegraphics[width=\linewidth]{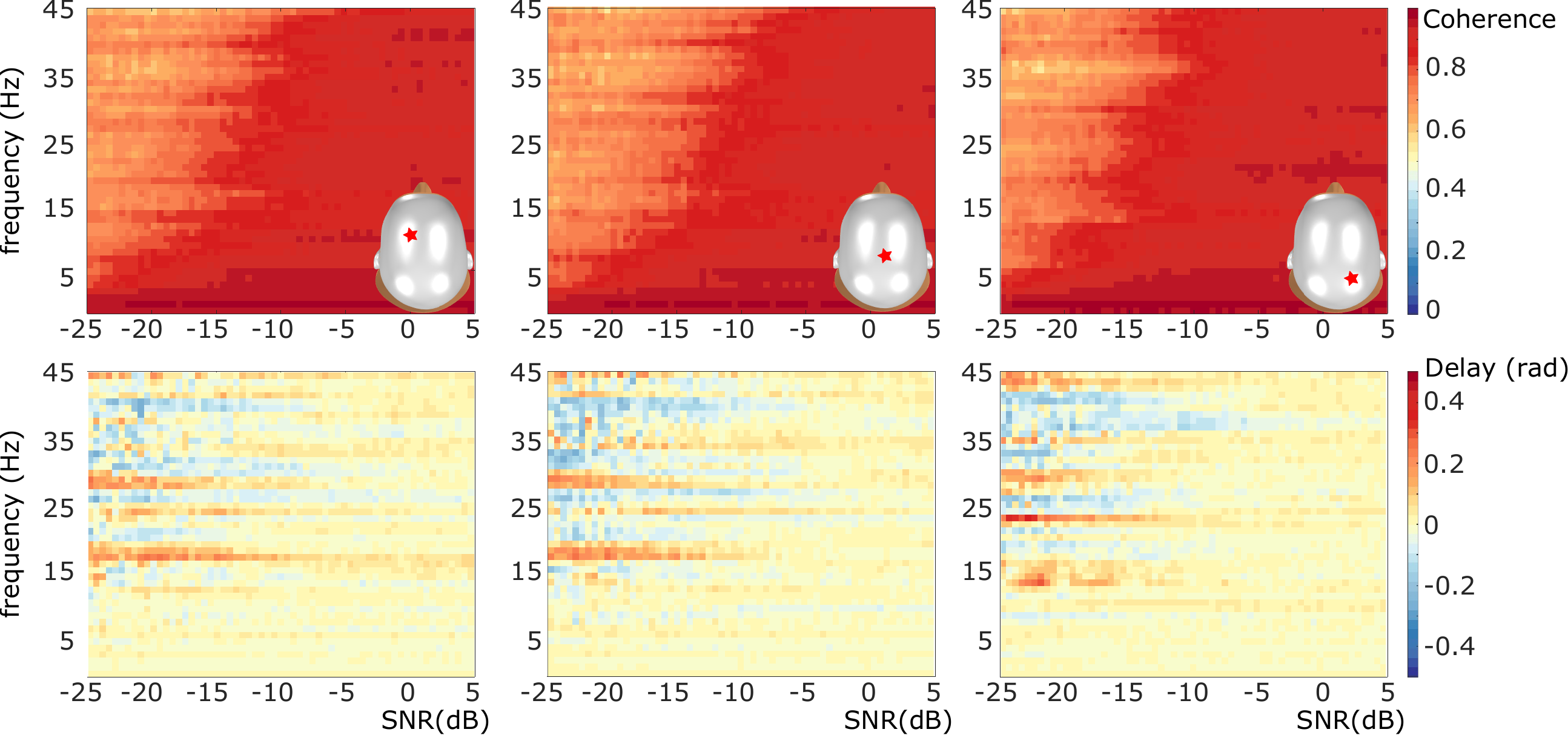} 
   \caption{Distortion produced by the SuBAR method on EEG epochs contaminated with ocular artifacts:  Same stipulations as in the caption of Figure~\ref{fig:artifactsCOHERENCE_EMG}.}
   \label{fig:artifactsCOHERENCE_EOG}
\end{figure}

To identify if the proposed algorithm preserves the natural frequency spectrum in denoised signals, we computed the spectral coherence and phase delays between the original clean EEG signals and the corrected data. Figures~\ref{fig:artifactsCOHERENCE_EMG}-\ref{fig:artifactsCOHERENCE_EOG} show that, even in presence of severe artifacts (SNR$<-10$ dB), the SuBAR method preserves the spectrum of EEG signals. A small amount of distortion on EEG spectrum is observed for frequencies larger than $15$Hz. In this band of frequencies, the correction of large artifacts introduced small absolute values of phase delay of about $0.1$ radians. Low frequencies ($f < 10$ Hz) were practically undistorted by the filtering method. Results clearly indicate that, in general, the proposed method preserves well the spectral components from clean brain signals and no significant phase delays are introduced by the filtering. 

% FIGURE 6

\subsubsection*{Correction of real contaminated data} 
\begin{figure*}[h!] 
   \centering%from left, bottom, right and top
   \includegraphics[width=\linewidth]{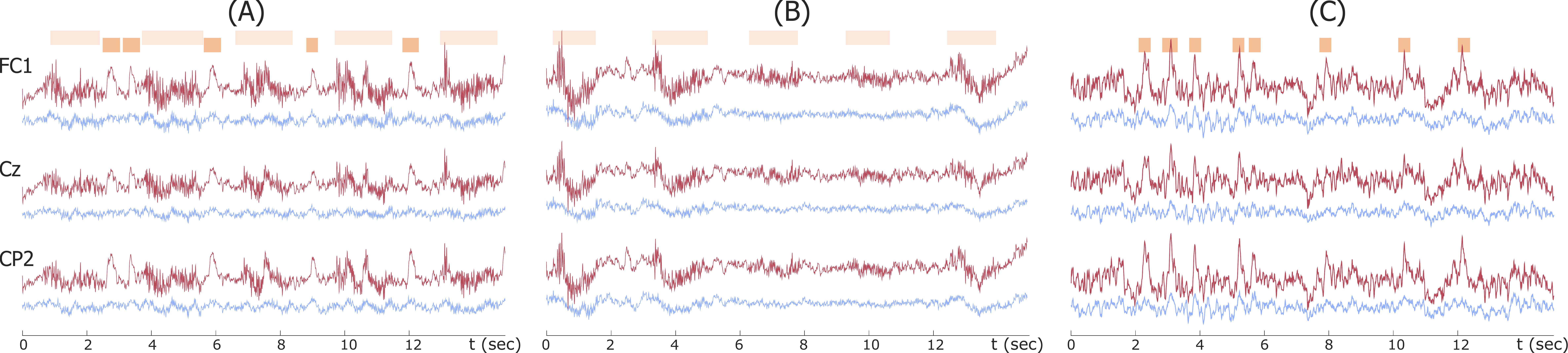} 
   \caption{Original EEG epochs contaminated with eye blinks and jaw clenching and the corresponding denoised signals:  (A) Epoch with mixed artifacts, (B) EEG segment containing muscular artifacts and (C) EEG epoch contaminated with eye movements. For illustration purposes only channels FC1, Cz and CP2 are shown as in Figs~\ref{fig:artifactsEMG}-\ref{fig:artifactsEOG}. Boxes in the top of the plots indicate the instances of jaw clenching and eye blinks.}
   \label{fig:artifactsRealEEG}
\end{figure*}

For illustration purposes, the SuBAR algorithm was applied on the EEG epochs from the second dataset, i. e. contaminated with real artifacts. Figure~\ref{fig:artifactsRealEEG}-(A) shows the performance of the SuBAR method on an EEG epoch that contains both eye blinks and muscular artifacts. Notice that, although different artifacts were present in the same segment, they were relatively well removed. In Figures~\ref{fig:artifactsRealEEG}-(B-C), we observe that, although small amounts of muscular activities remain in the denoised signals, our non-parametric algorithm corrects well the eye blinks from the contaminated data. Error measures could not be used here as we did not have a-priori information available of the clean brain signals.

% FIGURE 7

\subsubsection*{Computational complexity} For a given level of decomposition, the Maximal Overlap Wavelet Transform is computationally equivalent to other shift-invariant discrete wavelet transforms, and may be calculated with $\mathcal{O}(n\log{}n)$ computational complexity~\cite{complexityWavelets}. Although the wavelet thresholding and the surrogate-based algorithm have similar algorithmic complexity, the former remains highly advantageous from the computational point of view as surrogates are not generated. For the EEG segments analyzed here\footnote{Using Matlab on a 2.8GHz dual-core Intel Core i7 processor, 16GB of memory}, CPU times required by the WT method were, on average, hundred times faster than those required by the SuBAR filtering ($0.0075$~s \textit{vs} $1.37$~s, respectively). As expected, the CCA-EMD method requires larger CPU times to provide poorer performances ($14.48$~s).

% FIGURE 8

\section{Conclusion}
\label{conclusion}
In this work, a new data-driven method for automatic artifact removal in single-channel EEG was presented. The novelty of our proposal relies on the time-frequency analysis of surrogate data to identify and filter ocular and muscular artifacts embedded in single EEG channels. The efficacy of the algorithm was compared to wavelet thresholding, and the CCA combined with the EMD. Through artificially contaminated EEG signals, we demonstrate that the surrogate-based removal (SuBAR) algorithm outperforms the other techniques considered here for removing muscle and ocular artifacts from single EEG signals. Although large ocular artifacts are better removed by  wavelet thresholding, the SuBAR method yields, in general, a relative error 4 to 5 times smaller than the other considered artifact removal methods. Results show that the proposed algorithm preserves well the frequency spectrum of EEG in denoised signal (without phase delay). Furthermore, our method yields the smallest distortion of signals when applied to artifact-free EEG segments. Though it is not the aim of this study, we envisage that possible further optimizations can be obtained with other families of wavelets.

Most artifact reduction techniques require multivariate EEG data, or auxiliary referential signals (e.g. EOG or EMG)~\cite{uriguen2015}. Common artifact removal algorithms  generally requires appropriate spectral and topographical parameters for the detection of EEG artifacts (eye blinks, saccades, muscle activity)~\cite{uriguen2015, chen2016}. In contrast, results presented in this work suggest that our single-channel technique can be a good non-parametric filter for artifact removal in off-line environments with a reduced number of sensors. Computational profiling shows that the proposed method could be suitable for pseudo real-time environments, such as the mobile monitoring of cognitive or emotional states. The use of recent distributed signal processing algorithms might speed-up the algorithm for real-time implementations, or for the analysis of multi-channel EEG datasets~\cite{bertrand2015}. The proposed data-driven SuBAR method could be used in combination with other temporal EEG features (as those used in artifact detection~\cite{mognon2011}) to facilitate further signal processing of single-channel EEGs. 

Although the generation of surrogate time series and the corresponding wavelet decomposition of large EEG segments might be a bottleneck for the processing speed in on-line applications, its technical simplicity (it is fully data-driven) makes the SuBAR method highly operable in mobile environments, such as ambulatory healthcare systems~\cite{ehrenfeldBOOK}, where there are only a few EEG channels, or even a single channel is available (e.g. sleep stage scoring or anesthesia monitoring).  The practicality, accuracy and reliability of an adaptive filter based on ou method remain, however, to be explored. 

% FIGURE 9

%%%%%%%%%%%%%%%%%%%%%%%%%%%%%%%%%%%%%%%%%%%%%%%%%%%%%%%%%%%%%%%%%%%%%%%%%
\section*{Acknowledgments}
F. De Vico Fallani is supported by the French Agence Nationale de la Recherche (programme ANR-15-NEUC-0006-02), and X. Navarro-Sune is financially supported by Air Liquide Medical Systems S.A. 
%%%%%%%%%%%%%%%%%%%%%%%%%%%%%%%   THE BIBLIOGRAPHY  %%%%%%%%%%%%%%%%%%%%%%%%%%%%%%%

\bibliographystyle{ieeetr}
%\bibliography{bvi}

\end{document}